\documentclass[aps,prl,twocolumn,10pt,showpacs,superscriptaddress,floatfix]{revtex4-1}

\usepackage{amssymb}
\usepackage{graphicx,amsmath}
\usepackage{xcolor}
\usepackage{bm}
\usepackage{times}

\begin{document}

\title{Detection of weak microwave fields with an underdamped Josephson junction}

\begin{abstract}
We have constructed a microwave detector based on the voltage switching of an underdamped Josephson junction, that is positioned at a current antinode of a $\lambda/4$ coplanar waveguide resonator. By measuring the switching current and the transmission through a waveguide capacitively coupled to the resonator at different drive frequencies and temperatures we are able to fully characterize the system and assess its detection efficiency and sensitivity. Testing the detector by applying a classical microwave field with the strength of a single photon yielded a sensitivity parameter of 0.5 in qualitative agreement with theoretical calculations.

\end{abstract}

\date{\today}
\author{G.~Oelsner}\email{gregor.oelsner@leibniz-ipht.de} \affiliation{Leibniz Institute of Photonic Technology, P.O. Box 100239, D-07702 Jena, Germany}
\author{C.~K.~Andersen}\affiliation{Department of Physics and Astronomy, Aarhus University, DK-8000 Aarhus, Denmark}
\author{M.~Reh\'{a}k}\affiliation{Department of Experimental Physics, Comenius University, SK-84248 Bratislava, Slovakia}
\author{M.~Schmelz}\affiliation{Leibniz Institute of Photonic Technology, P.O. Box 100239, D-07702 Jena, Germany}
\author{S.~Anders}\affiliation{Leibniz Institute of Photonic Technology, P.O. Box 100239, D-07702 Jena, Germany}
\author{M.~Grajcar}\affiliation{Department of Experimental Physics, Comenius University, SK-84248 Bratislava, Slovakia}
\author{U.~H\"ubner}\affiliation{Leibniz Institute of Photonic Technology, P.O. Box 100239, D-07702 Jena, Germany}
\author{K.~M\o lmer}\affiliation{Department of Physics and Astronomy, Aarhus University, DK-8000 Aarhus, Denmark}
\author{E.~Il'ichev}\affiliation{Leibniz Institute of Photonic Technology, P.O. Box 100239, D-07702 Jena, Germany}
\affiliation{Novosibirsk State Technical University, 20 K. Marx Ave., 630092 Novosibirsk, Russia}

\maketitle
The light emission by single, microscopic quantum systems displays a number of non-classical features which have been exploited in fundamental investigations in quantum physics and which may result in applications in metrology, quantum communication and computing. Potential applications, however, would suffer from the rather weak coupling between atoms and single optical photons. This has stimulated efforts to study the same features with macroscopic artificial atoms. A particularly successful system relies on solid state superconducting circuits. Due to the Josephson non-linearity such circuits have an anharmonic excitation spectrum and may be restricted to an effective two-level systems which can interact resonantly with microwave fields. Besides the stronger coupling of superconducting circuits, an additional advantage is that they can be designed and fabricated on chip-scale, thereby allowing the integration in and scaling to larger systems with multiple components.

Essential quantum optical effects with superconducting qubits, such as vacuum Rabi splitting \cite{Wallraff2004}, resonance fluorescence of a single artificial atom \cite{Astafiev2010}, and single atom lasing \cite{Astafiev2007} have already been observed. Microwave fields can be amplified, detected and fully characterized in homodyne set-ups \cite{daSilva2010}. The effective coupling to transmission wave guides has made it possible to efficiently monitor the emitted radiation and verify the validity of the quantum trajectories of qubits conditioned on the detection record \cite{Murch2013, Ibarcq2016}, as well as to apply feedback and stabilize coherent superposition states of the qubit \cite{Vijay2012}.

Quantum optics benefits from high efficiency single photon detectors. It relies on the energy of the individual photons being sufficient to exploit the photoelectric effect and liberate an electron which can be amplified and detected \cite{Hadfield2009}. Transition edge sensors \cite{Foertsch2015} and superconducting nanowire single photon detectors \cite{Kahl2015,Natarajan2015} also require a sufficiently large energy of the incident photon to heat and thus modify the current through the detector. The energy of microwave photons is too low to allow detection by these methods, and for this energy range both controllable single photon sources and efficient single photon detectors are still under development. When working in the single photon regime, it is an obvious choice to use the resonant coupling to qubit systems. Indeed, the creation of single photons has been demonstrated by transferring the excitation from an excited qubit to a cavity \cite{Houck2007, Pechal2014} and to an emission line \cite{Peng2015}. Conversely, a qubit inside a microwave resonator modifies the reflection and transmission properties of the resonator and, hence, a resonator photon can be detected by transferring it first to a qubit excitation which is subsequently probed in a dispersive manner. For the detection of weak incident microwave fields, promising results have been obtained recently by the use of thermometry at superconductor$-$normal-metal$-$superconductor \cite{Govenius2014} and normal-metal$-$insulator$-$superconductor \cite{Gasparinetti2015} junctions as well as by the use of underdamped Josephson junction phase qubits \cite{Weides2011,Chen2011,Andersen2013}.

In this Letter, we consider the microwave field detection with a setup based on a current biased Josephson junction. These junctions can carry a lossless supercurrent up to a critical value $I_\mathrm{c}$ and they switch to a stable finite voltage state when this current is exceeded, or earlier, if a weak microwave field is applied to the device. The value of the output voltage can be several millivolts depending on the material combination used for the junction fabrication, which is easily detectable. The crucial part of the detector is an optimal coupling of the junction to the microwave field to achieve an efficient and sensitive field amplitude detection. For this purpose, a superconducting microwave resonator may be used. It is important to note that microwave detection by a current biased phase qubit is fundamentally different from both the qubit excitation by absorption of a single photon and from a conventional homodyne detection. The characterization of the device is crucial for the very definition of what is a measurement and what does it signify.

\begin{figure}[tb]
    \includegraphics[width=8 cm]{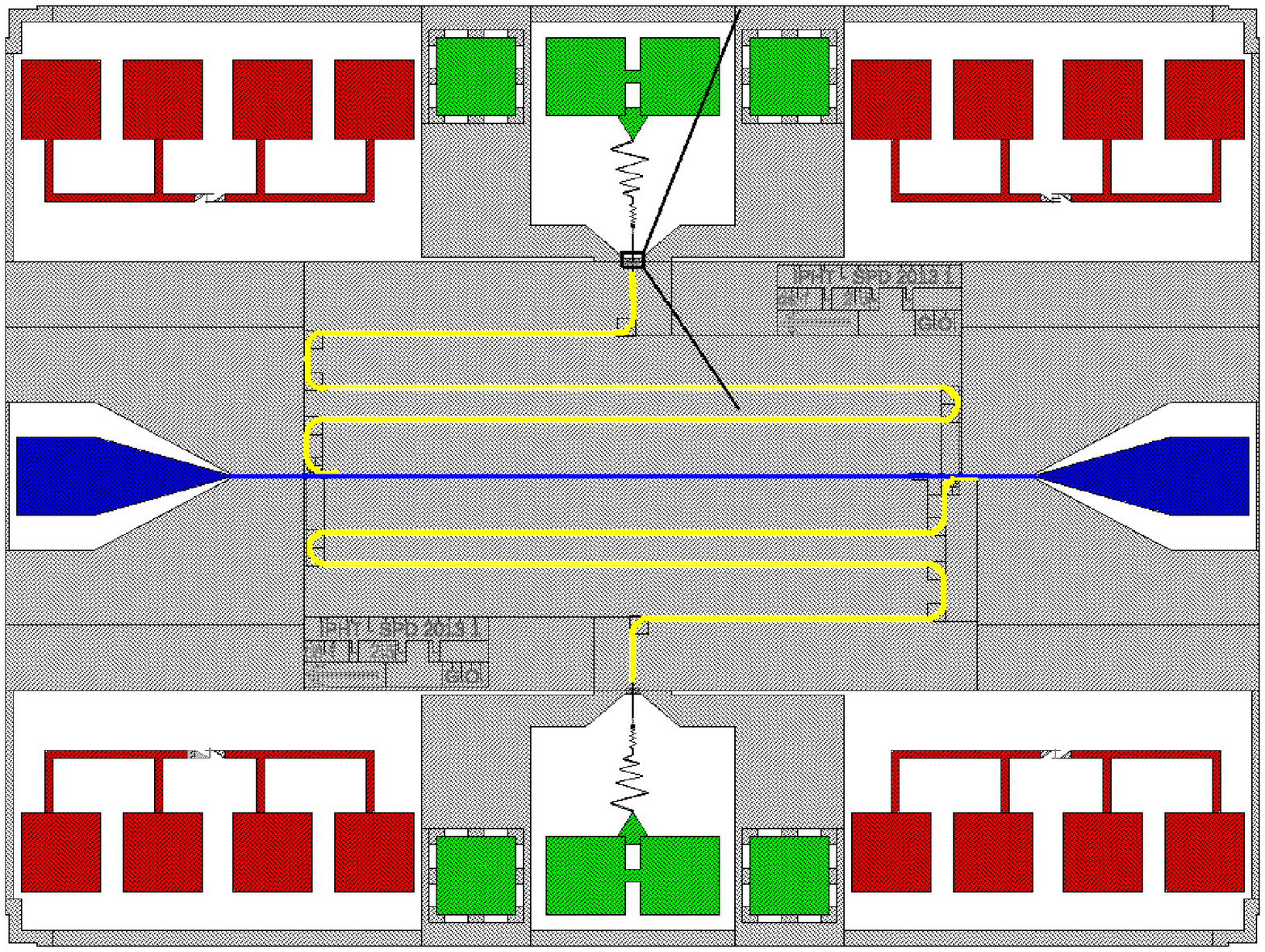}
    \begin{picture}(100,0)    
        \put(69.5,108){\includegraphics[width=3.2 cm]{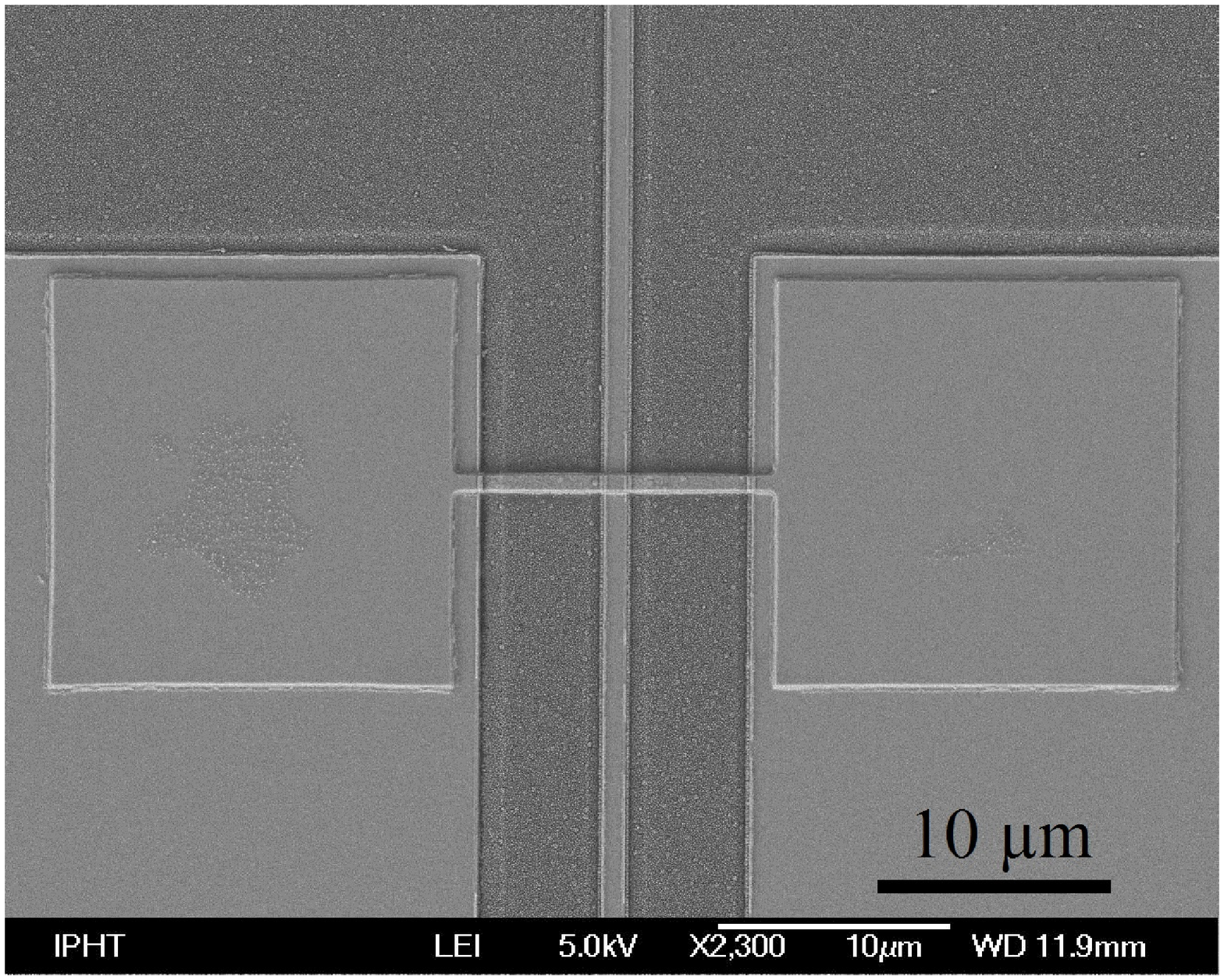}}
        \put(69.5,108){\line(0,1){73}}
        \put(69.5,181){\line(1,0){91}}
        \put(161,108){\line(0,1){73}}
        \put(69.5,108){\line(1,0){91}}
    \end{picture}
  \caption{(Color online). The $8 \times 6$~cm sized sample is displayed as a colored design picture consisting of a central conduction line (blue), capacitive coupled to two $\lambda/4$ resonators (yellow). The resonators are shunted in their current antinodes via Josephson junctions as shown in the insert SEM image. Each junction has four DC-connections (green) for current bias and voltage output. In the corners of the chip similar circuitry (red) are introduced to test and pre-calibrate the Josephson junctions. }\label{Fig:sample}
\end{figure}

In \cite{Andersen2014} we used a quantum trajectory treatment to analyze the performance of such a system comprised of a $\lambda/4$ coplanar waveguide (CPW) resonator shunted to ground via a Josephson junction, and in this Letter we demonstrate a prototype of such a device and report on its performance. We have designed and fabricated a superconducting circuit that consists of two $\lambda/4$ waveguides, both shunted to ground via a Josephson junction. The chip layout together with an SEM-image of one of the Josephson junctions is shown in Fig. \ref{Fig:sample}.
For the application of microwave signals and for the characterization of the system a transmission line is capacitively coupled to the resonators. The coupling capacitance is estimated to be $C_\mathrm{c} = 10$~pF by a full electromagnetic finite element simulation. The length of the resonators is chosen for a center frequency of the fundamental resonance at about $2.5$~GHz. A slight length difference is introduced to detune their resonant frequencies by about $60$~MHz to allow the individual characterization of each device \cite{Jerger2011}. The sample is fabricated on a silicon substrate using the cross-type Nb-AlOx-Nb fabrication process \cite{Anders2009} with a target critical current density of $200$~A/cm$^2$. The output (gap) voltage of the Josephson junctions takes a value of $2.8$~mV and no additional amplifiers are necessary for the detection.
The sample is placed in a dilution refrigerator with a base temperature of about $15$~mK. The external noise is reduced by magnetic and superconducting shielding together with heavy filtering of input and output signal lines. To estimate the influence of the external DC-lines on the microwave properties of the oscillators, we bonded only one junction to the external lines.

\begin{figure}[tb]
  \includegraphics[]{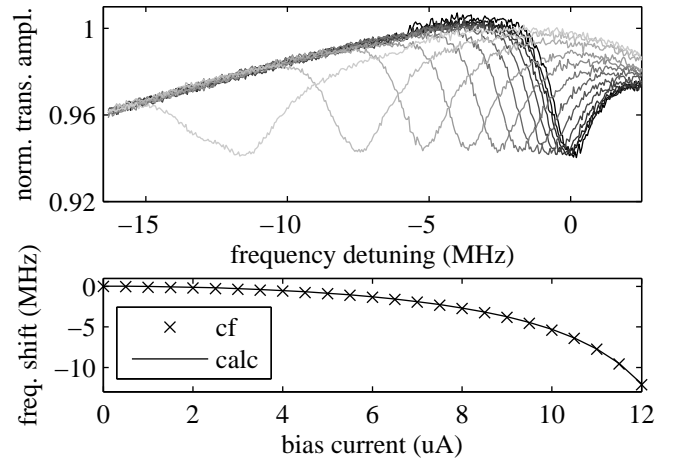}
  \caption{The upper panel shows the normalized transmission amplitude as a function of the probing frequency. From dark to light curves the DC-bias current at the junction is increased in steps of $1$~$\mu$A. The crosses in the lower panel show the center frequency (cf) shifts as a function of the bias current. They are found by fitting the transmission dips in the upper panel to Lorentzian line shapes. The measured frequency shifts are in excellent agreement with the analytical fit in Eq.\eqref{Eq:josephson_inductance} (solid line). }\label{Fig:josephson_inductance}
\end{figure}

With a first set of experiments we determined the parameters of the sub-systems. First, we measured the transmission of the central transmission line at different frequencies, and we found two resonances at $\omega_1/2\pi = 2.506$~GHz and $\omega_2/2\pi = 2.44$~GHz with similar quality factors of about 1000. Thus we conclude that no significant additional losses are introduced by contacting the junction. By applying a DC-bias current to the connected Josephson junction we are able to measure its resonance frequency $\omega_1$ shift to lower values, see the lower panel in Fig.~\ref{Fig:josephson_inductance}.

The Josephson inductance is a function of the bias current, $L_\mathrm{J} = \Phi_0 /(2 \pi \sqrt{I_\mathrm{c}^2 - I^2})$, where $\Phi_0$ is the magnetic flux quantum. Therefore, the total resonator inductance $L_\mathrm{t} = L_\mathrm{E} + L_\mathrm{J}$ also depends on $I$.  By making use of the lumped element representation of the oscillator we obtain an equation for its resonance frequency,
\begin{equation} \label{Eq:josephson_inductance}
\omega_1 = \frac{1}{\sqrt{C_\mathrm{E} \left( L_\mathrm{E} + L_\mathrm{J} \right)}}.
\end{equation}
By fitting this expression to the data we are able to extract the effective resonator capacitance $C_\mathrm{E} = 1.6$~pF, its inductance $L_\mathrm{E} = 2.5$~nH, and the critical current, $I_\mathrm{c} = 13.7$~$\mu$A. Knowing these parameters we can estimate the mean amplitude of the current $I_\mathrm{ph} \approx \sqrt{\hbar \omega_1 / L_\mathrm{E}} = 25$~nA produced by a single photon in the resonator. Note that vacuum fluctuations are of the same order of magnitude, and therefore it is important that we characterize the actual difference of the response of the device for the vacuum and weak microwave fields \cite{Andersen2014}.

\begin{figure}[tb]
  \includegraphics[]{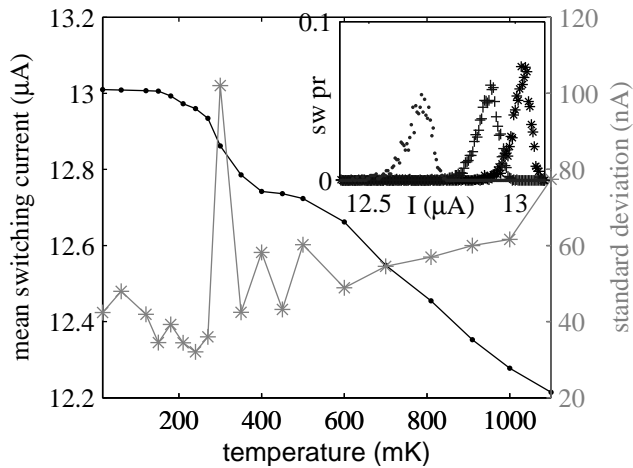}
  \caption{Mean switching current (black) and standard deviation (gray) of the switching current histograms at different temperatures. The insert shows examples of the histograms (switching probability over bias current) at temperatures 600, 300, and 14 mK marked by points, crosses and stars, respectively. }\label{Fig:temp_dep}
\end{figure}
In the next series of experiments we determined the switching current distributions under different experimental conditions by ramping up the current with a rate of $0.5$~$\mu$A/s and recording the values at which the junction switches to the finite voltage state. The extracted values for the mean switching current and their standard deviations $\sigma$ are plotted in Fig.~\ref{Fig:temp_dep} for different temperatures of the chip. At the lowest temperatures we found the largest values of the switching current $I_\mathrm{m}$ of about $13$~$\mu$A with a standard deviation $\sigma$ of $40$~nA. The measured switching currents are in qualitative agreement with the expected temperature dependence \cite{Oelsner2013}. In particular, we observe a crossover from a quantum to a thermally activated tunneling regime around $200$~mK, which is consistent with a normal resistance of $R_\mathrm{J} = 140$~$\Omega$ and a capacitance of $C_\mathrm{J} = 400$~fF \cite{Anders2009}, estimated from the junction size and geometry.

These experiments all served for characterization of the system parameters. Now we use the device to sense a weak microwave field. The application of a driving signal $V_\mathrm{in}\sin\omega_\mathrm{p} t$ yields a resonator driving amplitude of $\Omega_\mathrm{d} = C_\mathrm{c} V_\mathrm{in} V_0/2\hbar$. Here, $V_\mathrm{in}$ is the input voltage amplitude at the coupling capacitance and $V_0 = \sqrt{\hbar \omega_1 / C_\mathrm{E}}$ the zero point voltage of the resonator. In the experiments, carried out at 15 mK, we ramp the bias current, we observe the switching behavior, and record the switching current distributions as functions of the applied microwave frequency and amplitude.

\begin{figure}[tb]
  \includegraphics[width = 8.5 cm]{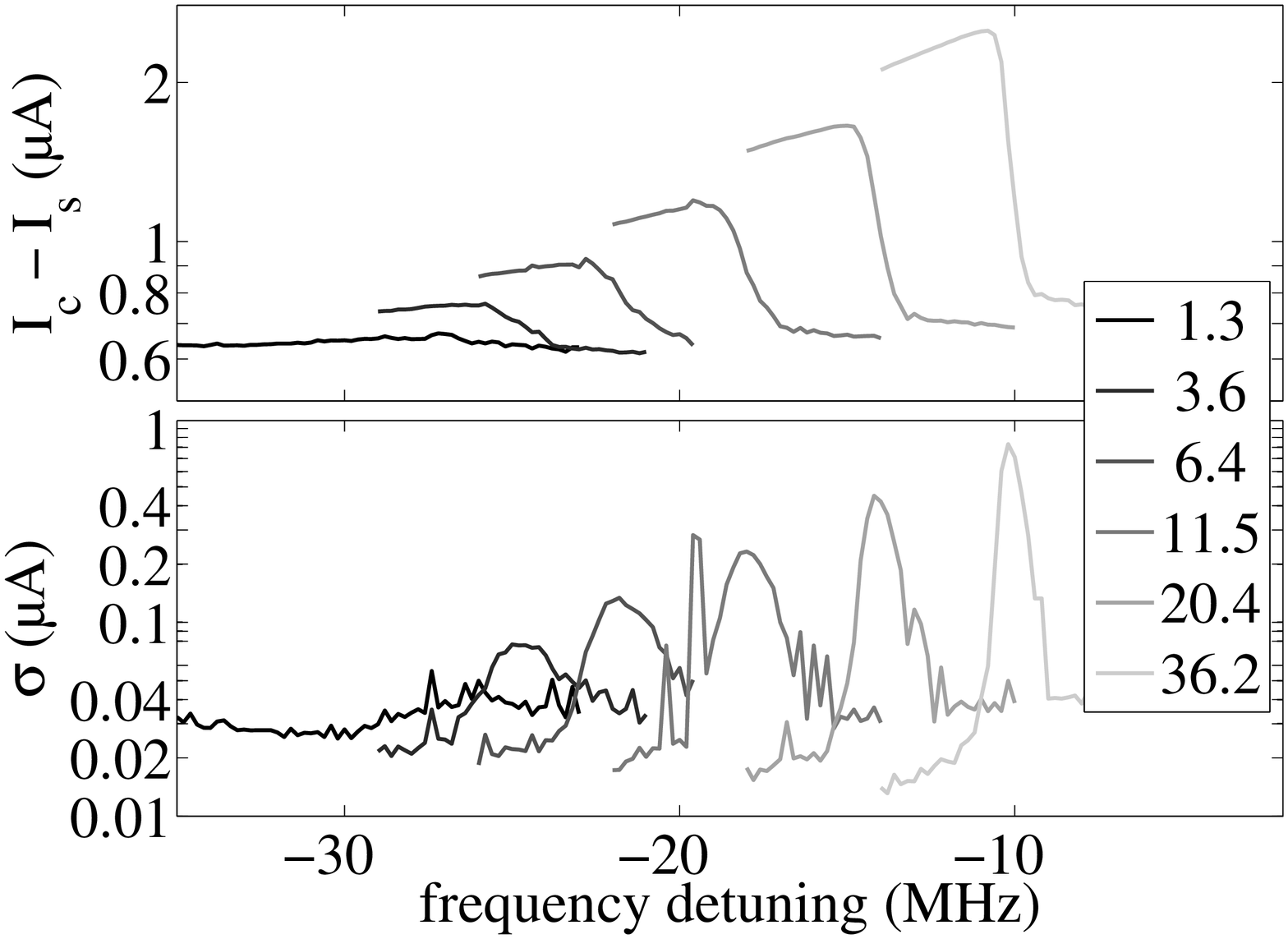}
  \caption{Influence of microwave signals to the switching current histograms. Difference between mean switching and critical current $I_\mathrm{c} - I_\mathrm{s}$ in the upper as well as the standard deviation $\sigma$ of the switching current in the lower picture measured for different driving frequencies. From the light to dark color the driving amplitude is decreased according to the legend in MHz.}\label{Fig:histogramshift}
\end{figure}

In Fig.~\ref{Fig:histogramshift}, the different curves display the frequency dependence of the mean switching current (upper panel) and its standard deviation (lower panel) for different values of the applied amplitude. The curves appear from the left to the right according to the increasing values of the amplitude indicated in the lower panel. The curves for different amplitudes all show a similar dependence on the frequency detuning.

The overall frequency dependence can be explained by the shift of the resonance curve with increasing bias current (see Fig.~\ref{Fig:josephson_inductance}). To detect a given amplitude of the rf-signal, the bias current must attain a specific value such that the additional rf-current suffices to switch the junction to the finite voltage state. Due to the non-linear Josephson inductance, this bias current shifts the resonator frequency, and hence the optimal input coupling is achieved at the shifted resonance condition of the resonator. When decreasing the frequency for the large amplitude driving, we obtain a standard deviation with a minimum, that lies below the one of the undisturbed junction. This is because the strong driving allows excitation directly into the continuum that is represented by the voltage state in addition to the switching by tunneling \cite{Andersen2013}.

\begin{figure}[tb]
  \includegraphics[width = 8.5 cm]{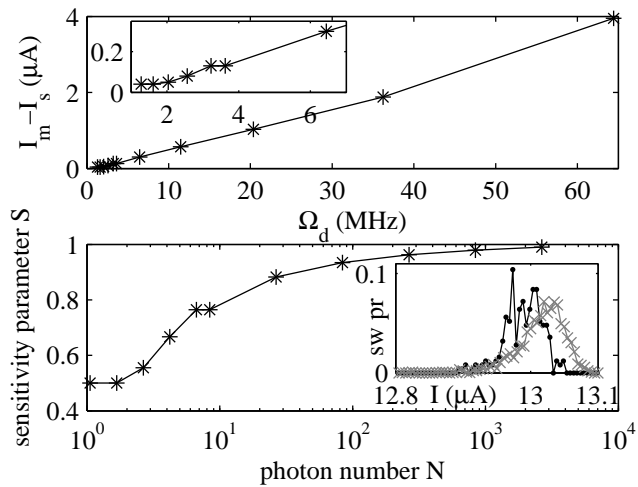}
  \caption{(Upper plot) Maximal shift of the mean switching current at different driving amplitudes $\Omega_d$. The insert shows a zoom of the region close to zero driving. (Lower plot) Reconstructed sensitivity of our detector as extracted maximal shift of the histograms divided by the shift plus the standard deviation of the switching current of the undisturbed junction. The inset shows the undisturbed histogram (grey crosses) compared to the maximally shifted histograms for the lowest applied driving power (black dots). They have a Hellinger distance \cite{Spehner2014} of 0.51. The shifted histogram shows a characteristic peak structure in qualitative agreement with theoretical calculations.
 }\label{Fig:sensitivity}
\end{figure}

Finally, from the same experimental data, we extract the dependence of the maximal shifts of the mean switching currents as function of the driving amplitude and corresponding photon number in the resonator (see Fig.~\ref{Fig:sensitivity}). For the device to work as a detector of single photons, the design needs to be tailored such that the sensitivity is maximal at the frequency of the photons. The bandwidth of the detector is then set by the linewidth appearing in Fig.~\ref{Fig:histogramshift}.

The upper panel of Fig.~\ref{Fig:sensitivity} displays a linear dependence of the maximal shift of the mean switching currents on the driving amplitude. This indicates that the switching to the voltage state that leads to the detection is, indeed, caused by the modulation of the junction potential by the rf-current in the coupled resonator and junction system \cite{Andersen2014} and not just by resonance between the applied microwave and the qubit.

The inset in the lower plot of Fig.~\ref{Fig:sensitivity} shows switching current distributions for the undisturbed and the weakly driven junction. To estimate whether the detector is illuminated with a given field strength or not by a single measurement of the switching current, one merely identifies which of the two candidate switching current probability distributions is largest at the value measured. The accomplishment of that procedure, averaged over the possible outcomes, is quantified statistically, e.g., by the Hellinger distance \cite{Spehner2014} between the distributions. For Gaussian distributions, this distance is given analytically by their widths and separation, and for simplicity, we shall characterize our (non-Gaussian) distributions by a width $\sigma$ and by their mean values $I_\mathrm{m}$ and $I_\mathrm{s}$. The amplitude signal-to-noise ratio can then be defined as $SNR = (I_\mathrm{m} - I_\mathrm{s})/\sigma$.  This motivates the assignment of the sensitivity parameter, $S = SNR/(SNR +1)$, denoting how well we can distinguish whether a drive is applied. By plotting $S$ as a function of the mean photon number $N = 4 \Omega_\mathrm{d}^2/\kappa^2$ \cite{Shevchenko2014} in the lower panel of Fig.~\ref{Fig:sensitivity} we find an SNR of 1 and a value of $S=0.5$ for a drive strength corresponding to a single microwave photon in the resonator. Here, a current amplitude of 25~nA is expected.

Using the theory of \cite{Andersen2014} we can calculate the effective junction potential and subsequently obtain the switching rate
\cite{Caldeira1981,Andersen2013} and the theoretical switching current distributions for different values of the drive strength. This qualitatively reproduces the experimental data, including the characteristic spiked structure, seen in the inset of Fig.~\ref{Fig:sensitivity}, which we ascribe to the coupling between the resonator and the current-biased Josephson junction. Our calculation, however, underestimates the width ($\sigma = 10$ nA) and thus yield a slightly higher sensitivity parameter of 0.65 in the single photon regime. A simple theoretical consideration suggests that a device with a smaller critical current will have a higher sensitivity.
The detection also works for stronger microwave signals and after a calibration at a fixed frequency detuning appearing in Fig.~\ref{Fig:histogramshift}, we can deduce the classical mean photon number in a signal injected on the detector.

In conclusion, we developed the prototype of a device that may be used for the determination of classical field amplitudes in the microwave domain. Our device achieves a sensitivity parameter of 0.5 in the low photon limit. This value is in agreement with the expected signal level of single photons, and it may be improved by altering the design. For example the width of the switching current distributions may be decreased, and for values as achieved in \cite{Oelsner2013} a sensitivity parameter of the order of unity can be expected at the single photon level. Further, the input coupling of the cavity may be optimized and the resonator can be matched to fit the impedance of standard transmission lines and thus avoid reflections. Based on the work presented here, such optimized devices are currently under development in our laboratory and will be made subject for future investigations.

\begin{acknowledgments}
The research leading to these results has received funding from the European
Community Seventh Framework Programme (FP7/2007-2013) under Grant No. 270843
(iQIT). C.K.A. and K.M. acknowledge the financial support from the Villum Foundation. M.G. and M.R. acknowledge partial support from the Slovak Research and Development Agency under the Contract Nos. APVV-0808-12 and APVV-0088-12.
\end{acknowledgments}

\end{document}